\documentclass[aps,prl,reprint,groupedscriptaddress]{revtex4-1}
\usepackage{graphics}
\usepackage{amsmath}
\usepackage{mathrsfs}
\usepackage{amssymb}
\usepackage{bm}
\usepackage{multirow}
\begin{document}

\title{Antiferromagnetic resonance excited by oscillating electric currents}

\author{Volker Sluka}
\affiliation{Department of Physics, New York University, New York, NY 10003, USA}

\date{\today}

\begin{abstract}
In antiferromagnetic materials the order parameter exhibits resonant modes at  frequencies that can be in the terahertz range, making them interesting components for spintronic devices. Here, it is shown that antiferromagnetic resonance can be excited using the inverse spin-Hall effect in a system consisting of an antiferromagnetic insulator coupled to a normal-metal waveguide. The time-dependent interplay between spin-torque, ac spin-accumulation and magnetic degrees of freedom is studied. It is found that the dynamics of the antiferromagnet effects the frequency-dependent conductivity of the normal metal. Further, it is shown that in antiferromagnetic insulators, the resonant excitation by ac spin-currents can be orders of magnitude more efficient than excitation by the current-induced Oersted field.
\end{abstract}


\maketitle
\begin{figure}[h!]
\includegraphics{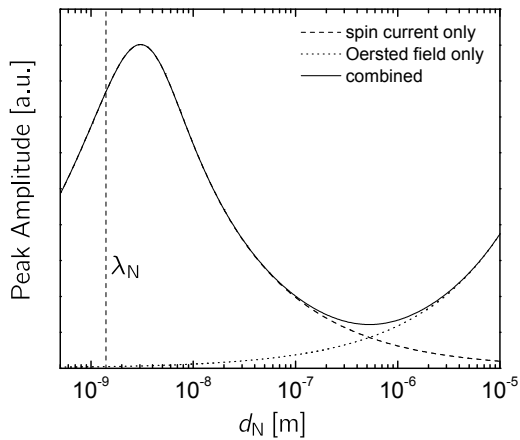}
\caption{\label{fig-5} AFM precession amplitude at resonance under the condition of constant power injection into the wave guide. The initial increase of the amplitude results from the build-up of spin-accumulation, hence $\lambda_{\mathrm{N}}$ is the relevant length scale. At higher NM thickness, the total current and thus the influence of the Oersted field increase. This trend is however limited by the skin depth (not included in the calculation), which is estimated to be hundreds of naometers for the frequencies.}
\end{figure}

Spin-transfer torque (STT) and giant magnetoresistance \cite{Baib1988,Barn1990,Slonc1996,Berg1996,Myer1999,Kati2000} form the foundation of spintronics, together with more recent additions such as spin-orbit torques, whose prominent manifestations are the spin-Hall and inverse spin-Hall effects (SHE/ISHE) \cite{Dyak1971,Vale2006,Taka2008}. The latter, besides their usefulness for device applications, have also developed into standard experimental methods of spin current generation and detection. The interplay of spin-currents and ferromagnetic materials continue to be at the core of developments towards magnetic random access memory (STT-MRAM), sensors, and radio-frequency components.\\
\indent Parallel to these efforts, antiferromagnetic materials have recently been considered as active components in spintronic applications. So far, most theoretical studies show that, similarly to ferromagnets, the order parameter of an antiferromagnet can be manipulated by spin-transfer torque \cite{Gomo2010,Gomo2012, Chen2014,Chen2015,Dani2015} or by optical and magnetic pulses \cite{Wien2012,Gomo2016,Higu2016}, and other excitation mechanisms have been proposed \cite{Seki2016}. Insulating antiferromagnets are also intensively studied in conjunction with
\begin{figure}[h!]
\includegraphics{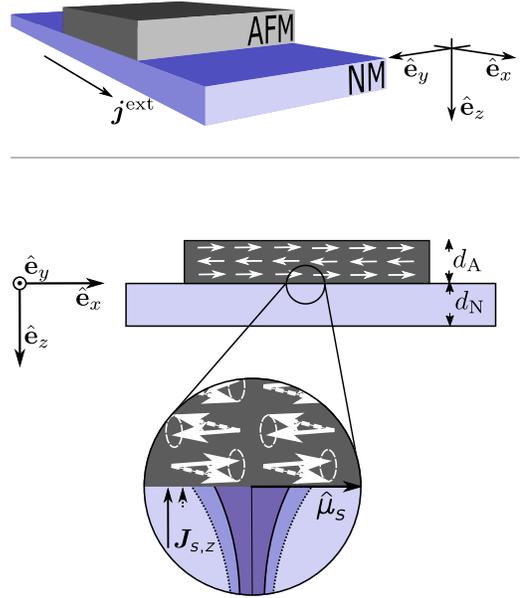}
\caption{\label{fig-1} Schematic of the structure. An antiferromagnetic insulator is coupled to a normal metal layer with pronounced spin-orbit coupling, which is subjected to a rapidly oscillating charge current $\bm{j}^{\mathrm{ext}}$ in the $x$-direction. The order parameter of the antiferromagnet is excited by the spin-Hall effect in a sample geometry that is typically used in ferromagnetic resonance experiments. The inset illustrates the processes taking place near the interface between both layers: as a result of the oscillating spin-current, an ac spin-accumulation with amplitude $\hat{\mu}_{s}$ builds up, as represented by the continuous lines extending from the interface into the normal metal layer. When driven near antiferromagnetic resonance (dashed lines in the normal metal and dashed arrows in the antiferromagnet), the spin-accumulation amplitude changes, leading to changes in the total spin current $\bm{J}_{s,z}$ across the interface.}
\end{figure}
spin transport \cite{Hahn2014, Wang2014, Mori2015, Hung2017}.\\
\indent In the context of practical applications, antiferromagnets exhibit a number of advantages compared to ferromagnets \cite{Baltz2017}: First, their compensated magnetic moments lack any stray fields, which have been identified as potentially limiting the performance of ferromagnet-based STT-MRAM. In addition, antiferromagnet dynamics can be significantly faster than in ferromagnets --- their response can be at THz frequencies. Besides the conventional generation of STT through perpendicular-to-plane injection of polarized dc or --- in the case of STT ferromagnetic resonance \cite{Sank2008} --- ac  charge currents, the SHE can be employed for the same purpose. In latter case, a nonmagnetic metal (NM) with pronounced spin-orbit coupling is placed adjacent to the magnetically ordered layer. A charge current parallel to the plane of the NM layer will result in spin-current, whose spatial direction is set by the vector product of the charge current and the spin-polarization direction \cite{Taka2008}. Studies exploiting the SHE in this particular geometry have so far mostly been performed on ferromagnetic and ferrimagnetic materials \cite{Liu2011, Zhou2013, Jiao2013, Chi2014,Schr2015, Chen2016}, and few on antiferromagnetic layers \cite{Manc2017}.\\
\indent In this work, an experiment is proposed that takes advantage of the SHE in order to study antiferromagnets in a sample geometry similar to that used to do ferromagnetic resonance spectroscopy, where the role of the Oersted field is replaced by an oscillating spin-current. Resonant excitation of antiferromagnetic insulators by spin-orbit torques can be of significantly higher efficiency compared to that by Oersted fields. In deriving the frequency-dependent waveguide conductivity it will be shown that the resonant precession of the antiferromagnetic order parameter effects the ac electrical transmission properties of the waveguide. The proposed method can be used to investigate in principle any antiferromagnetic insulator thin film, including $\mathrm{MnF_{2}}$ \cite{Jaco1961,Hagi1999,Ross2013} and $\mathrm{NiO}$ \cite{Hutc1972, Mach2017, Hou2017, Mori2015}.\\
\indent The considered system consists of an antiferromagnetic insulator (AFM) with two symmetric sublattice magnetizations which is assumed to exhibit a uniaxial anisotropy. The AFM is coupled to a normal metal (NM) thin film with strong spin-orbit interaction that serves as a wave-guide for an rapidly oscillating ac charge current.\\
\indent The coordinate system is depicted in Fig.\,\ref{fig-1}: The $x$-axis is the direction of current propagation and coincides with the direction of uniaxial anisotropy in the AFM. The $z$-axis is the direction perpendicular to the NM/AFM interface, which is assumed to lie in the $x$-$y$ plane ($z=0$). Denoting the thicknesses of NM and AFM films by $d_{\mathrm{N}}$ and $d_{\mathrm{A}}$, respectively, the NM film occupies the space $[0,d_{N}]$ and the AFM is located in $[-d_{A},0]$. An externally imposed oscillating charge current density of frequency $\omega=2\pi f$, $\bm{j}^{\mathrm{ext}}(t)=\hat{\bm{e}}_{x}j_{0}\cos (\omega t)$, flows along the $x$-direction, where $\hat{\bm{e}}_{i}$, $i\in \left\{x,y,z \right\}$ denotes the unit vector along the $i^{\mathrm{th}}$ direction. That current arises from a time-dependent electric field $\bm{E}^{\mathrm{ext}}(t)=E(t)\hat{\bm{e}}_{x}$, so that $\bm{j}^{\mathrm{ext}}(\omega)=\sigma (\omega) \bm{E}^{\mathrm{ext}}(\omega)$, where $\sigma (\omega)$ is the frequency-dependent conductivity of the NM.\\
\indent As the goal is to investigate processes in the scale of hundreds of GHz and higher, the frequency dependence of the conductivity will be included in the calculation. In the frequency domain, and to first order in the excitation,
\begin{equation} \bm{\mu}_{s}(z,\omega)=\sigma_{\mathrm{sf}}(\omega)D(\omega)\frac{\partial^{2}}{\partial z^{2}}  \bm{\mu}_{s}(z,\omega) \label{eqn1}
\end{equation}
for the spin-accumulation $\bm{\mu}_{s}$ in the normal metal, with
$\sigma_{\mathrm{sf}}(\omega):=i/(\omega+i/\tau_{\mathrm{sf}})$,
and
$D(\omega):=\frac{1}{3}\sigma_{\mathrm{tr}}(\omega)v_{\mathrm{F}}^{2}$,
where $\sigma_{\mathrm{tr}}(\omega):=i/(\omega+i/\tau_{\mathrm{tr}})$, $D(0)$ coinciding with the diffusion constant in the stationary spin-diffusion equation.  $\tau_{\mathrm{tr}}$ and $\tau_{\mathrm{sf}}$ denote the transport and spin-flip relaxation times, respectively, and $v_{\mathrm{F}}$ is the Fermi velocity. With $\kappa(\omega)=[\sigma_{\mathrm{sf}}(\omega)D(\omega)]^{-\frac{1}{2}}$, the general solution of Eq. (\ref{eqn1}) is
\begin{equation}
\bm{\mu}_{s}(z,\omega)=\bm{A}(\omega)\exp[-\kappa(\omega)z]+\bm{B}(\omega)\exp[\kappa(\omega)z].
\end{equation} 
The frequency-dependent spin current along the $z$-direction, caused by the gradients in the spin-accumulation and the external electric field via the ISHE is then
\begin{equation}
\bm{J}_{s,z}(z,\omega)=-\sigma(\omega)\theta_{\mathrm{SH}}E(\omega)\hat{\bm{e}}_{y}-\frac{\sigma(\omega)}{2e}\frac{\partial \bm{\mu}_{s}(z,\omega)}{\partial z},
\end{equation}
where $e>0$ is the elementary charge and $\theta_{\mathrm{SH}}$ denotes the Spin-Hall angle.
\begin{figure}[h!]
\includegraphics{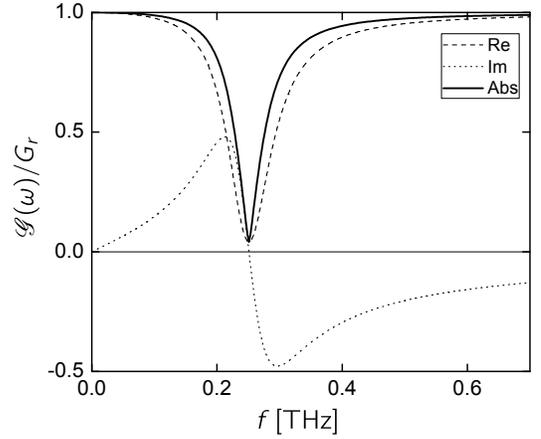}
\caption{\label{fig-3} The effective conductance $\mathscr{G}$, as defined in Eq.\,(\ref{EffCond}) further below, determines the influence of the AFM magnetization dynamics on the spin-accumulation at the NM/AFM interface. This example is based on a $\mathrm{Pt}/\mathrm{MnF_{2}}$ bilayer with the sample parameters in Table \ref{tab-1}, and $d_{\mathrm{A}}= 5\, \mathrm{nm}$. The spin-mixing conductance is set to  $1.9 \times 10^{14}\, \mathrm{\Omega^{-1} m^{-2}}$. The dashed, dotted and solid lines correspond to the real part, imaginary part and the absolute value of $\mathscr{G}$, respectively. The peak position corresponds to the AFM resonance frequency.}
\end{figure}
The requirement that the above spin current vanishes at the lower boundary of the normal metal, $\bm{J}_{s,z}(-d_{N},\omega)=0$, allows to express  spin-accumulation with $\bm{A}(\omega)+\bm{B}(\omega)=\bm{\mu}_{s}(0,\omega)$ as the single parameter:
\begin{align}
\cosh[\kappa(\omega) d_{N}] &\bm{\mu}_{s}(z,\omega)=\cosh[\kappa(\omega)(d_{N}-z)] \bm{\mu}_{s}(0,\omega)\nonumber\\
&-2e\kappa^{-1}(\omega)\sinh[\kappa(\omega) z]\hat{\bm{e}}_{y}\theta_{\mathrm{SH}}E(\omega),
\end{align}
Accordingly, the spin current at the NM/AFM interface is
\begin{align}
\bm{J}_{s,z}&(0,\omega)=[\sigma(\omega)/(2e)]\kappa(\omega)\tanh[\kappa(\omega)d_{N}]\bm{\mu}_{s}(0,\omega)\nonumber\\
&-2\sigma(\omega)
\frac{\sinh^{2}[\kappa(\omega)d_{N}/2]}{\cosh[\kappa(\omega)d_{N}]}\hat{\bm{e}}_{y}\theta_{\mathrm{SH}}E(\omega).
\end{align}
\indent We now consider the spin currents in the AFM.
There are two contributions \cite{Brat2012,Chi2014,Gomo2010}, one related to the spin-transfer torque,
\begin{equation}
\bm{J}^{\mathrm{STT}}_{s,z}(t)=\frac{G_{r}}{2e}\sum_{i}\bm{m}_{i}(t)\times [\bm{m}_{i}(t) \times \bm{\mu}_{s}(0,t)]\label{eqn2},
\end{equation}
and the one arising from spin pumping
\begin{equation}
\bm{J}^{\mathrm{SP}}_{s,z}(t)=\frac{\hbar G_{r}}{2e}\sum_{i}\bm{m}_{i}(t)\times \frac{\mathrm{d}}{\mathrm{d}t}\bm{m}_{i}(t)\label{eqn3}.
\end{equation}
In Eqs. (\ref{eqn2}) and (\ref{eqn3}), $\bm{m}_{i}$ ($i \in \left\{1,2\right\}$) denote the two sublattice magnetization unit vectors, and for simplicity it is assumed that the spin-mixing conductance $G_{r}$ is real. This assumption can be made, since its imaginary part is usually orders of magnitude smaller than its real part \cite{Chen2014}. The goal is to solve the continuity condition at the NM/AFM interface,
\begin{equation}
\bm{J}_{s,z}(0,t)=\bm{J}^{\mathrm{STT}}_{s,z}(t)+
\bm{J}^{\mathrm{SP}}_{s,z}(t)\label{eqn4}
\end{equation}
for $ \bm{\mu}_{s}(0,t)$, while considering that the magnetization dynamics of the sublattices is governed by two coupled Landau-Lifshitz-Gilbert equations \cite{Chi2014,Gomo2010},
\begin{align}
\frac{\mathrm{d}\bm{m}_{i}}{\mathrm{d}t}=& -\gamma \bm{m}_{i}\times \bm{H}_{i}+\alpha \bm{m}_{i}\times \frac{\mathrm{d}\bm{m}_{i}}{\mathrm{d}t}\nonumber\\
&+\gamma G\bm{m}_{i}\times[\bm{m}_{i}\times \bm{\mu}_{s}(0,t)].
\end{align}
\begin{figure*}
\includegraphics{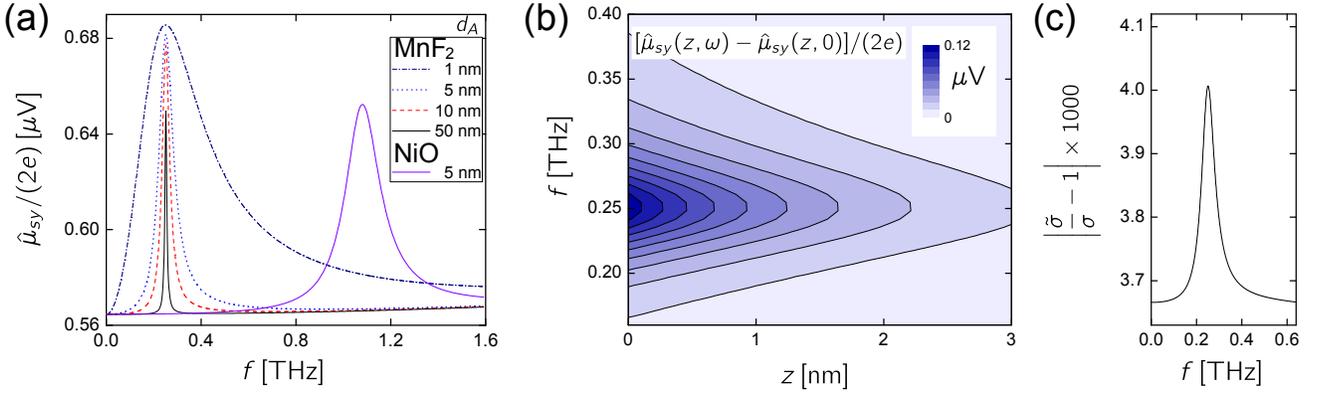}
\caption{\label{fig-4} Time domain spin-voltage amplitudes $\hat{\mu}_{sy}/(2e)$ as a function of the excitation frequency $\omega=2\pi f$, computed for the example systems $\mathrm{Pt}/\mathrm{MnF_{2}}$ and $\mathrm{Pt}/\mathrm{NiO}$ with the material parameters displayed in Table \ref{tab-1}. For the spin-mixing conductance $G_{r}$, a value of $1.9 \times 10^{14}\, \mathrm{\Omega^{-1} m^{-2}}$ is assumed, the normal metal thickness is set to $d_{\mathrm{N}}= 10\, \mathrm{nm}$ and $j_{0}=10^{6}\,\mathrm{A/cm^{2}}$.  In (a), the line profiles are shown for various thicknesses of the AFM layer. The line positions correspond to the AFM mode with frequency $\omega = \gamma \sqrt{ (h_{k}-h_{p}) (h_{k} - 2 \lambda)}$. The frequency shift between the two materials is largely associated with the order of magnitude higher exchange coupling in $\mathrm{NiO}$. Higher AFM layer thicknesses lead to narrow linewidths, which is a consequence of the spin-pumping contribution to the damping of the AFM oscillations. The resonant driving of the AFM by the spin-current leads to increased spin pumping, which creates a shift in the local spin-accumulation in the normal metal. The spatial and frequency dependence of the induced change in spin-accumulation is shown in (b) for $\mathrm{Pt}/\mathrm{MnF_{2}}$ and $d_{\mathrm{A}}= 5 \, \mathrm{nm}$. The changes in the spin currents associated with the resonant precession of the AFM magnetic moments cause changes in the net charge current, as a consequence of the SHE. The resulting effect on the mean current density is captured by an effective conductance shown in panel (c).}
\end{figure*}
The exchange coupling between the lattice magnetizations is mediated by the effective fields $\mu_{0}M_{s}\bm{H}_{i}=-\nabla_{\bm{m}_{i}}\mathscr{E}$, derived from the energy density $\mathscr{E}$ with $(\mu_{0}M_{s})^{-1}\mathscr{E}=-\lambda\bm{m}_{1}\cdot\bm{m}_{2}-\bm{h}_{\mathrm{ext}}\cdot\sum\bm{m}_{i}-\frac{1}{2}
h_{k}\sum(\hat{\bm{e}}_{x}\cdot\bm{m}_{i})^{2}-\frac{1}{2}
h_{p}\sum(\hat{\bm{e}}_{z}\cdot\bm{m}_{i})^{2}$. Here, $M_{s}$ is the saturation magnetization of each sublattice. Furthermore, $\gamma$ is the gyromagnetic ratio, while $h_{k}$ and $h_{p}$ denote the strength of two uniaxial anisotropies to include biaxial cases such as $\mathrm{NiO}$. The axes of $h_{k}$ and $h_{p}$ are oriented along the $x$- and $z$- direction, respectively. $\lambda$ determines the strength of the exchange interactions between the sublattice magnetic moments, and $G=\hbar G_{r}/(4e^{2}\mu_{0}M_{s}d_{A})$ is the scaled (real) spin-mixing conductance. The external magnetic field is assumed to consist only of the time dependent Oersted field $h_{\mathrm{Oe}}$ generated by the oscillating charge current, $\bm{h}_{\mathrm{ext}}=h_{\mathrm{Oe}}\hat{\bm{e}}_{y}$. Finally, the effective damping $\alpha$ is the sum of the Gilbert damping $\alpha_{0}$ and a contribution $\gamma \hbar G$, accounting for spin-pumping.\\
\indent The AFM resonance frequency can be as high as $0.1$-$1\,\mathrm{THz}$, \textit{i.e.}, it is comparable to $\tau^{-1}_{\mathrm{sf}}$ and $\tau^{-1}_{\mathrm{tr}}$. Thus, the spin-accumulation in the NM and the magnetization dynamics in the AFM take place on the same time scale, which requires a self-consistent solution of Eq. (\ref{eqn4}). This is in contrast to previous work considering similar sample geometries \cite{Chi2014,Chen2016}, where the dynamical time scales of the magnetic system and the conductor could be separated, and the time dependence only entered through the boundary conditions. To this end, it is helpful to note that the interest is in small angle precession of the magnetizations. It is therefore sufficient to expand to first order in the deviations of the sublattice magnetizations from the ground state, where the lattice magnetizations are largely oriented along the anisotropy axis, \textit{i.e.} $\bm{m}_{i}\approx(-1)^{i-1}\hat{\bm{e}}_{x}+\delta m_{i,y}\hat{\bm{e}}_{y}+\delta m_{i,z}\hat{\bm{e}}_{z}$. The dynamic parts of the lattice magnetizations, the spin-accumulation and the Oersted field can be considered as small and of first order in $E^{\mathrm{ext}}$. Then, the linearized equation of motion for the magnetizations reads
\begin{equation}
\frac{\mathrm{d}}{\mathrm{d}t}\delta \bm{m}(t)
=\mathscr{M}\, \delta \bm{m}(t)+
\mathscr{N} h_{\mathrm{Oe}}(t)
+ \, \mathscr{A} \, \bm{p}(t),\label{eqn4b}
\end{equation}\\
with $\delta \bm{m}(t):=[\delta m_{1y}(t),\delta m_{1z}(t),\delta m_{2y}(t),\delta m_{2z}(t)]^{T}$ and $\bm{p}(t)=[0,\mu_{sy}(0,t),\mu_{sz}(0,t)]^{T}$. Furthermore, in Eq. (\ref{eqn4b}),
\small
\begin{align}
&-(1+\alpha^{2})\gamma^{-1}\mathscr{M}:=\nonumber\\
&\begin{pmatrix}
\alpha (h_{k}-\lambda) & h_{k}-h_{p}-\lambda & -\alpha \lambda & -\lambda\\
-h_{k}+\lambda & \alpha (h_{k}-h_{p}-\lambda) & \lambda & -\alpha \lambda \\
 -\alpha \lambda & \lambda & \alpha (h_{k}-\lambda) & -h_{k}+h_{p}+\lambda\\
 -\lambda & -\alpha \lambda & h_{k}-\lambda & \alpha (h_{k}-h_{p}-\lambda)\nonumber
\end{pmatrix},
\end{align}
\normalsize
\begin{align}
\mathscr{A}:=-\frac{\gamma}{1+\alpha^{2}}G
\begin{pmatrix}
0& 1 & -\alpha \\
0& \alpha & 1  \\
0& 1 &\alpha\\
0& -\alpha & 1\nonumber
\end{pmatrix},
\end{align}
and $\mathscr{N}:=-\frac{\gamma}{1+\alpha^{2}}(-\alpha,1,-\alpha,-1)^{T}$. Then, to first order, the expressions (\ref{eqn2}) and (\ref{eqn3}) can be written as
\begin{equation}
\bm{J}^{\mathrm{STT}}_{s,z}(\omega)=-\frac{G_{r}}{e}\bm{p}(\omega),\label{eqn5}
\end{equation}
and
\begin{table*}[t]
\caption{\label{tab-1} Material parameters used in the calculations}
\begin{ruledtabular}
\begin{tabular}{ccccccc|ccccc}
\multicolumn{7}{c|}{Antiferromagnet parameters \cite{Ross2013,Hutc1972,Sato2010,Mach2017}}&\multicolumn{5}{c}{Normal metal parameters \cite{Weil2013,Obst2014}} \\
\,& $M_{s}$  & $h_{k}$ &$h_{p}$ &  $\lambda$ & $\alpha_{0}$ & $\gamma $&$\sigma(0)$  & $\tau_{\mathrm{tr}}$ & $\tau_{\mathrm{sf}}$ & $\lambda_{\mathrm{N}}$ & $\theta_{\mathrm{SH}} $ \\
\,& $[\mathrm{Am^{-1}}]$  & $[\mathrm{Am^{-1}}]$& $[\mathrm{Am^{-1}}]$ & $[\mathrm{Am^{-1}}]$ & \, & $ [\mathrm{mA^{-1}s^{-1}}]$&$[\mathrm{\Omega^{-1}  m^{-1}}]$  & $[\mathrm{fs}]$ & $[\mathrm{fs}]$ & $[\mathrm{nm}]$ & $\, $ \\
\hline
$\mathrm{MnF_{2}}$ &$4.77\times 10^{4}$ & $6.76 \times 10^{5}$ &$0$ & $-4.22 \times 10^{7}$ & $1.0\times 10^{-3}$ & $2.08 \times 10^{5}$ &\multirow{2}{*}{$2.44 \times 10^{6}$}&\multirow{2}{*}{$18$}&\multirow{2}{*}{$10$}&\multirow{2}{*}{$1.4$}&\multirow{2}{*}{0.12}\\
$\mathrm{NiO}$ &$5.60\times 10^{5}$ & $9.79 \times 10^{3}$ & $-5.71 \times 10^{5}$ & $-6.81 \times 10^{8}$ & $1.0\times 10^{-3}$ & $2.41 \times 10^{5}$ \\
\end{tabular}
\end{ruledtabular}
\end{table*}
\begin{align}
\frac{e}{\hbar G_{r}}\bm{J}^{\mathrm{SP}}_{s,z}(\omega)=&
  \mathscr{L}[-i\omega -\mathscr{M}]^{-1}[\mathscr{N}\, h_{\mathrm{Oe}}(\omega)
+ \mathscr{A} \, \bm{p}(\omega)]\nonumber\\ 
&+\frac{\gamma}{1+\alpha^{2}}[h_{\mathrm{Oe}}(\omega)\hat{\bm{e}}_{y}+\alpha G\, \bm{p}(\omega)] ,\label{eqn6}
\end{align}
where
\begin{align}
&\mathscr{L}:=-\frac{1}{2} \frac{\gamma}{1+\alpha^{2}}\times \nonumber\\
&\times \begin{pmatrix}
0&0 &0 & 0\\
h_{k}-2\lambda & -\alpha (h_{k}-h_{p}) &h_{k}-2\lambda & \alpha (h_{k}-h_{p}) \\
 \alpha h_{k} &h_{k}-h_{p}-2\lambda& -\alpha h_{k} &h_{k}-h_{p}-2\lambda
\end{pmatrix}.\nonumber
\end{align}
With Eqs. (\ref{eqn5}) and (\ref{eqn6}), Eq. (\ref{eqn4}) can be solved for $\bm{\mu}_{s}(0,\omega)$. In the symmetric sample orientation considered here, only the $y$-component of the spin-accumulation is non-vanishing:
\begin{align}
\frac{\mu_{sy}(0,\omega)}{e\theta_{\mathrm{SH}}E(\omega)}=\left[ \frac{\cosh(d_{N}\kappa)\mathscr{G}(\omega)}{2\sigma\sinh^{2}(d_{N}\kappa/2)}+\coth\left(\frac{d_{N}\kappa
}{2}\right)\frac{\kappa}{2}
  \right]^{-1}\label{eqn7}
\end{align}
with the effective conductance
\begin{widetext}
\begin{align}
\mathscr{G}(\omega)=G_{r}
\frac{ 
   \gamma^{2} (h_{k} - 2 \lambda) (h_{k}-h_{p} + i G \hbar \omega) + 
   \alpha \gamma \omega [  G \hbar \omega - i (2h_{k}-h_{p} - 2\lambda)] -(1 + \alpha^{2}) \omega^{2} }{ \gamma^{2} (h_{k}-h_{p}) (h_{k} - 2 \lambda) -  i \alpha \gamma (2h_{k}-h_{p} -2 \lambda) \omega - (1 + \alpha^{2}) \omega^{2}}\label{EffCond}
\end{align}
\end{widetext} 
The quantity $\mathscr{G}(\omega)$ contains all the properties of the AFM and establishes the link between the two subsystems of wave-guide and magnet. An example for $\mathscr{G}(\omega)$, computed for $\mathrm{MnF_{2}}$, is shown in Fig. \ref{fig-3}. The line-shaped feature reflects the increased spin pumping activity when the AFM is driven near resonance. With $\mathscr{G}(\omega)$, it is possible to compute the self-consistent spin-accumulation resulting from the interplay with the AFM. Examples are shown in Fig.\,\ref{fig-4}.\\
\indent The influence of the Oersted field generated by the oscillating current is much smaller than that of the SHE, and therefore its contribution to the result (\ref{eqn7}), has been omitted. In Fig.\,\ref{fig-5}, the two excitation mechanisms are compared as a function of waveguide thickness under the experimentally relevant condition of constant power injection. Starting at low $d_{\mathrm{N}}$, the AFM precession amplitude first increases, as the spin-accumulation builds up on the length scale of the spin diffusion length $\lambda_{\mathrm{N}}$. As $d_{\mathrm{N}}$ is increased further, the current density and thus the influence of the spin-transfer torque decrease, until the Oersted field starts to dominate at $d_{\mathrm{N}}\approx 0.5\, \mathrm{\mu m}$. At this length scale, however, the skin depth is expected to limit further increase of the Oersted field's influence.\\
\indent The spin-accumulation imbalance induced by the AFM dynamics influences the charge transport through the NM layer via the SHE. The waveguide thickness-averaged charge current density and the electric field are related by the effective electric conductivity $\tilde{\sigma}$ with
\begin{align}
\frac{\tilde{\sigma}}{\sigma}-1=4\theta^{2}_{\mathrm{SH}}\frac{\sinh(\frac{\kappa d_{\mathrm{N}}}{2})\left[\kappa \sigma \sinh(\frac{\kappa d_{\mathrm{N}}}{2})+\mathscr{G}\cosh(\frac{\kappa d_{\mathrm{N}}}{2})\right]}{d_{\mathrm{N}}\kappa \left[\kappa \sigma \sinh(\kappa d_{\mathrm{N}})+2\,\mathscr{G}\cosh(d_{\mathrm{N}}\kappa)\right]}.\label{eqn8}
\end{align}
The above change in the conductivity is shown in Fig. \ref{fig-4}(c). The resonant dynamics of the AFM is transferred via the ISHE to the charge flow through the spin-Hall metal, resulting in an effective conductivity that exhibits a peak at the resonance frequency of the AFM. The effect is caused by the magnetization dynamics; it takes place on top of the spin-accumulation background set by the so-called Spin-Hall magnetoresistance \cite{Naka2013,Chen2013}. The size of the feature is small due to the averaging over the waveguide and the fact that it is of second order in $\theta_{\mathrm{SH}}$. However, transport effects of comparable size have been measured at GHz frequencies \cite{Lotz2014}, and modules extending network analyzer operation to the THz range are available.\\
\indent In summary, the analysis here considers the time-dependent interplay between the AFM spin dynamics and the spin-transport in the waveguide on an equal, self-consistent footing. The proposed method provides an efficient mechanism to excite and study spin-waves in normal metal / antiferromagnet thin film systems. From a practical point of view, ac-spin currents generated by spin-orbit torques provide a new way to study antiferromagnetic resonance in insulators and thus characterize their magnetic interactions, including their magnetic anisotropy, exchange and damping.\\

\indent I would like to thank A. D. Kent for fruitful discussions. This work was supported in part by NSF-DMR-1610416.\\

\end{document}